\documentclass[aip,pof,reprint]{revtex4-1}

\usepackage{amsmath}
\usepackage{graphicx}
\usepackage{upgreek}

\newcommand{\be}{\begin{equation}}
\newcommand{\ee}{\end{equation}}

\newcommand{\bd}[1]{\mathbf{#1}}

\newcommand{\rmi}{\mathrm{i}}
\newcommand{\rmd}{\mathrm{d}}
\newcommand{\rme}{\mathrm{e}}

\newcommand{\half}{{\textstyle \frac1{2}}}
\newcommand{\sixth}{{\textstyle \frac1{6}}}

\begin{document}

\title{Electrostrictive fluid pressure from a laser beam}
\author{Simen \AA.\ Ellingsen} 
\author{Iver Brevik} 
\affiliation{Fluids Engineering Division, Department of Energy and Process Engineering, Norwegian University of Science and Technology, N-7491 Trondheim, Norway}

\begin{abstract}
Recent times have seen surge of research activity on systems combining fluid mechanics and electromagnetic fields. In radiation optics, whenever information about the distribution of pressure in a dielectric fluid is required, the contribution from electrostriction becomes important. In the present paper we calculate how the local pressure varies with position and time when a laser beam is imposed in a uniform fluid. A Gaussian intensity profile of arbitrary time dependence is assumed for the beam, and general results are derived in this case. For demonstration we analyze two different cases: first, that the beam is imposed suddenly (mathematically in the form of a step function); secondly, that the beam is switched on in a soft way. In both cases, simple analytical expressions for the pressure distribution are found. 
\end{abstract}
\maketitle

\section{Introduction}

Over the last decade, considerable experimental research has been conducted into the interplay of optics and fluid mechanics. The field of optofluidics, the application of microfluidic flows for optical purposes, is a topic in its infancy and is seen by many as holding very considerable technological promise \cite{monat07}. Appreciable progress in optical manipulation of microflows has also been achieved; some examples are found in Refs.~\onlinecite{garnier03,baroud07,brasselet08,verneuil09,delville09,wunenburger11}. These impressive experimental developments create the need also for improved theoretical understanding of liquid-laser interactions. As a step towards this end we have investigated herein the time-dependent electrostrictive pressure due to a laser beam propagating through a liquid.

The electrostriction force density is a local force density acting within all dielectric fluids and solids subjected to an electric field, proportional to gradient of the field intensity. The force acts towards higher field strength, thus causing a local pressure increase in the center of a laser beam propagating through the medium.
From a general viewpoint, when including also solid elastic media, the electrostrictive susceptibility is strongly related to the elasto-optic susceptibility as they both represent the interaction of electric fields and an elastic deformation in the medium. A very general account of media of this kind can be found in Nelson's book \cite{nelson79}, and a comprehensive statistical mechanical theory is provided by Rasaiah, Stell and co-workers\cite{rasaiah81,rasaiah81,carnie82,rasaiah82}. When restricting ourselves to electrostrictive phenomena in fluids we may mention, for instance, the early paper of Brueckner and Jorna \cite{brueckner66} in which it is discussed how the electrostrictive coupling between laser light and the medium can be important for rapidly growing instabilities, and hence may lead to large density fluctuations. Another work dealing with electrostriction in fluids is the series of papers by Zanchini and Baretta \cite{zanchini94, barletta94,barletta94b}; here a discussion is given of the role of electrostriction for the attractive force between capacitor plates completely immersed in a dielectric fluid.

A work of particular interest in our context is the microgravity experiment of Zimmerli {\it et al.}\ on electrostriction in a fluid under extraterrestrial conditions  \cite{zimmerli99}. This work is actually a generalization of the classic electrostriction experiment of Hakim and Higham \cite{hakim62} (The Hakim-Higham experiment is discussed more closely in Ref.~\onlinecite{brevik79}, section 2.3.). The last-mentioned authors measured the pressure between closely spaced electrodes immersed in a nonpolar liquid (carbon tetrachloride and n-hexane). 
The essential new element in  the experiment of Zimmerli {\it et al.}\ was to examine electrostriction in a liquid, sulfur hexafluoride (SF$_6$) near the liquid-vapor critical point, $T_c=319$ K, where the compressibility was seven orders of magnitude higher than in the liquids studied by Hakim and Higham. 
Because of this compressibility, even a moderate electric field produced an easily measurable density change in the nonpolar fluid. 
To avoid disturbances from stratification gravity effects on Earth, the experiment was carried out in  outer space, in the Space Shuttle Columbia in 1994. For the application of electrodynamic theory to fluids, the experiment is important.

However, generally speaking the occurrence of an electrostrictive pressure term in a dielectric fluid has for the most part received only moderate attention in practical situations. The reason for this is quite evident: the electrostrictive force density is generally written as a gradient, and the total electrostrictive force on a dielectric specimen, which may be written as an exterior surface integral, gives no contribution to the force on the specimen as a whole if the surroundings are non-dielectric. Under usual physical conditions, the electrostrictive effect therefore only becomes of importance in considerations of {\it local} pressure distributions in matter, accessible experimentally usually via optical methods.

Although deformations and flows generated by electromagnetic fields can generally be described without inclusion of electrostriction, this may not be so for their \emph{stability}. This point was emphasized by one of us already some time ago \cite{brevik81} and we return to it in the following few examples. The deformations of fluids by lasers, under static conditions as well as under time-dependent conditions, are therefore adequately described in terms of Maxwell's stress tensor, without electrostriction included at all. See, for instance, some of the Bordeaux papers in Refs.~\onlinecite{brasselet08,delville09,wunenburger11} and also some related theoretical papers in Refs.~\onlinecite{hallanger05,birkeland08}. For full stability considerations, however, electrostriction could play an important role. 

Electrostriction can moreover be made useful, for example in laser-induced thermal acoustics (LITA) which has been made a tool for measuring a range of quantities including the speed of sound in a medium\cite{cummings94,cummings95}. The electrostrictive force, being able to effectuate local changes in density, can also locally alter a medium's refractive index, giving rise to non-linear optical response even at moderate field strength\cite{huang04}, resulting in such effects as anomalous Stokes gain\cite{brueckner66}, and optical self-focusing\cite{self-focusingBook}. 

A few examples may serve to illustrate the typical role of electrostriction.
Firstly, it is noteworthy that the electrostrictive force is of the same order of magnitude as the usual electromagnetic force density
 $-(\varepsilon_0/2)E^2\boldsymbol \nabla \varepsilon$ acting in regions where  the permittivity $\varepsilon$ varies with position, typically in dielectric boundary layers.
  (We write the constitutive relations as ${ \bf D}=\varepsilon_0 \varepsilon \bf E$, ${\bf B}=\mu_0\mu \bf H$, so that $\varepsilon$ and $\mu $ become nondimensional.)
   Consider for definiteness the usual textbook situation in which two metal plates are partially immersed in a dielectric liquid (the situation is discussed, e.g.\ in section 2.2 of Ref.~\onlinecite{brevik79}). When there is an electric field between the plates, the
   liquid is known to rise to a height $h$ that can be easily calculated. The main physical principle here is the balance between the lifting surface force at the free
   surface and the gravitational force. A nontrivial point  is, however, the effect of the electrostrictive force: its presence  is necessary to secure that the excess electric
    pressure in the liquid is strong enough to make it possible for the liquid to rise as a coherent whole\cite{brevik81}. The electrostrictive effect thus serves to stabilize the
    system - it is intimately related to thermodynamic stability -  but it does not play a role for the magnitude of the height $h$.

The observability of electrostrictive
     phenomena depends on one single physical variable in the
     system, namely the velocity of sound. In the classic
     experiment of Ashkin and Dziedzic \cite{ashkin73}, our second example,
     a narrow beam of light was sent through a
     dielectric liquid and an elevation of the free surface (in the
     order of 1 $\upmu$m) was observed. The mean time between
     molecular collisions in such a  liquid is of order 1 ps, and sound
       needs only a time of  about 7 ns to traverse the cross section of the beam  (waist radius  $w_0= 4.5~\mu$m).  This means that the elastic pressure has had   sufficient time to
     build itself up long before  experimental effects become visible.  As shown from a detailed calculation in Fig. 9 in Ref.~\onlinecite{brevik79}, the maximum dynamic electrostrictive pressure occurred at times $t \sim $3 ns after the sudden onset of the pulse at $t=0$ and died out after  8-10  ns. By contrast, the hydrodynamical motion of the free surface needed hundreds of nanoseconds to develop. For practical purposes it is thus legitimate to introduce the
      concept of a hydrodynamical pressure $p$, and to assume  that the
     electromagnetic forces start to act simultaneously throughout
     the liquid.

A third example of essentially the same kind is
     furnished by the radiation pressure experiment of Zhang and Chang from
     1988 \cite{zhang88}: the authors observed the
     radiation-induced deformation of the surface of a
     micrometer-droplet illuminated by a laser beam. Again, the
     electrostrictive pressure in the interior of the droplet is
     counterbalanced by an elastic pressure having had sufficient
     time to build itself up. The deformation of the droplet can thus
     be calculated as though there were no electrostrictive effects at
     all. The theory of this experiment has been worked out in
     Refs.~\onlinecite{lai89,brevik99}.

Under special circumstances it becomes possible to make the electrostrictive effect discernible in force measurements even without relying upon optical methods. The Goetz-Zahn experiment belongs to this category \cite{goetz55,goetz58,zahn62}. The main principle of this experiment was to apply a high frequency electric field $E=E_0\cos \omega t$ between the two plates of a condenser filled with a liquid (polar or nonpolar), and to measure the appropriate harmonic component of the force between the plates. Strong fields were necessary for this purpose, of the order of $E=10^6~$V/m.  The reason why the electrostrictive part of the force was detected, was that the liquid was in a state of mechanical non-equilibrium; the fields were oscillating so quickly that the elastic counter pressure had insufficient time to build. To propagate a distance of 1 cm, sound needs a time of about 10 $\upmu$s, which is more than the period $2\pi/\omega \sim 1~\mu$s of the oscillations. This case was discussed in detailed in Ref.~\onlinecite{brevik79}, section 2.4.

In the following  we will calculate the distribution of the electrostrictive pressure in a liquid under typical experimental conditions. The practical utility of such an undertaking lies in the increasing need to have detailed information about the stress distribution in soft matter in the presence of lasers. In view of the increasing interest in liquid-laser interactions at a microscopically detailed level, such a calculation is timely.
We consider for demonstration two different time evolutions of the laser beam; the case in which the pulse is switched on and off abruptly, and the case of a softly peaked pulse with exponential tail, while our final result is valid for an arbitrary time evolution and can be of use, for example, in applications within laser-induced acoustics. In both cases relatively simple expressions are obtained for the pressure distribution in the liquid as a function of distance from the beam axis and time [corrections due to beam divergence are ignored]. A natural time and length scale for the problem are provided by the beam width and the velocity of sound. Our final results in the two cases are Eqs.~\eqref{pstep} and \eqref{Ppulse}, in which the pressure distribution is given as function of time and radius rescaled with respect to these scales.

\subsection{Basic formalism}

\begin{figure}[htb]
  \begin{center}
    \includegraphics[width=3.4in]{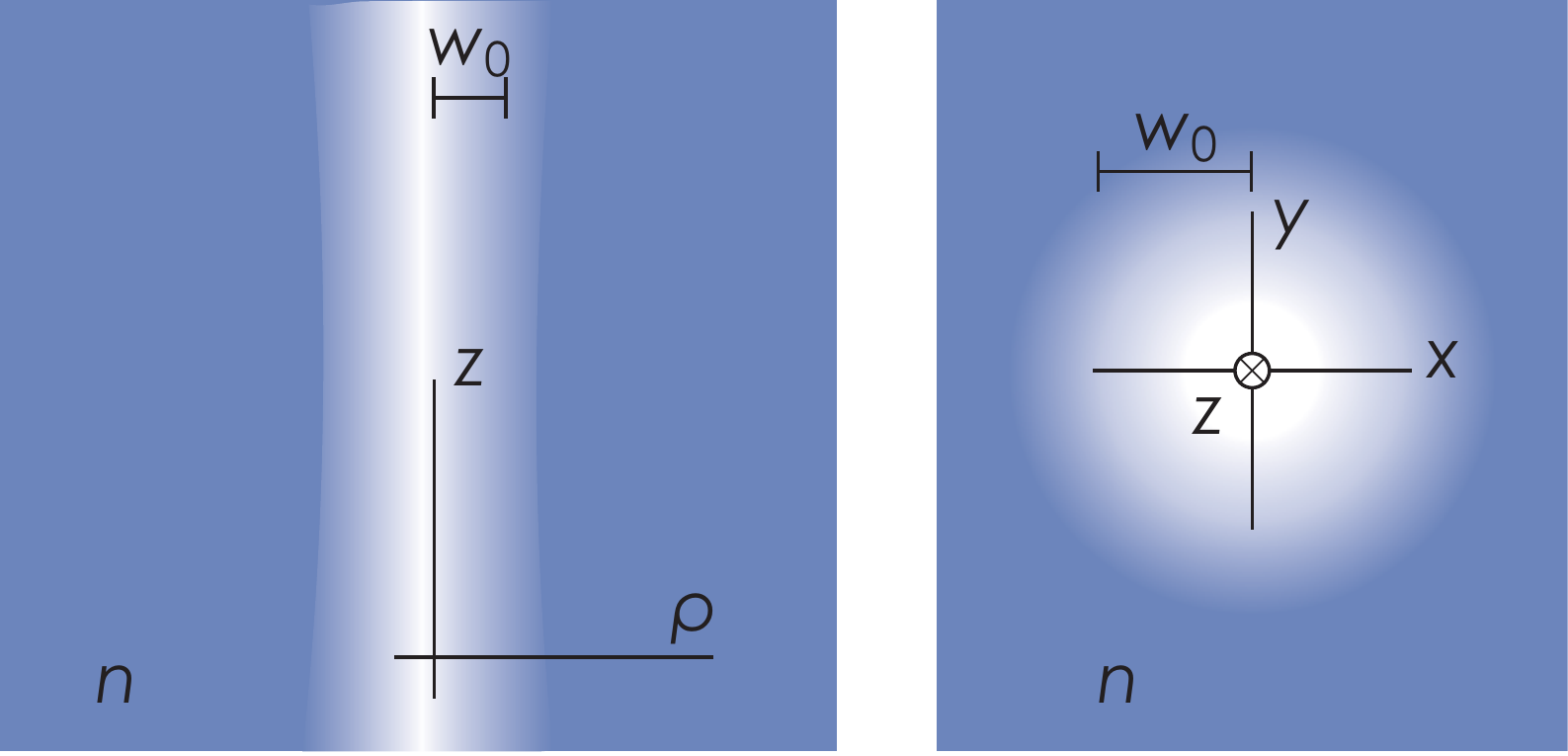}
    \caption{The situation considered: A Gaussian laser beam of width $w_0\gg \lambda$ propagates through a medium of relative permittivity $\varepsilon$ ($\lambda$: wavelength). The laser intensity varies in time according to a function $T(t)$.}
    \label{fig:geometry}
  \end{center}
\end{figure}

We regard a laser beam propagating along the $z$ direction in a homogeneous liquid, with beam intensity described by a Gaussian so that the squared electric field in cylindrical coordinates $\bd{r}=(\rho,\theta,z)$,
\be
  E^2(\bd{r},t) = E^2(\bd{r}) T(t)
\ee
with (c.f.\ e.g.\ Ref.~\onlinecite{brevik79} and chapter 16 of Ref.~\onlinecite{siegman86})
\be
  E^2(\bd{r}) = \frac{2 P}{\pi \varepsilon_0 n c w^2(z)} \exp\left[-\frac{2\rho^2}{w^2(z)}\right].
\ee
The situation is shown in figure \ref{fig:geometry}. Here, $P$ is the total incident laser power, $n$ is the liquid's index of refraction, and because of inevitable beam divergence its radius is $z$ dependent and of the form
\be
  w(z) = w_0 \sqrt{1+z^2/l_R^2}
\ee where $w_0$ is the waist radius and $l_R=\pi w_0^2/\lambda$ is
the Rayleigh length with $\lambda$ the wavelength of the laser
light in the liquid, and $c$ is the velocity of light in vacuum. In the following we shall assume the beam width $w_0$ to be much greater than the laser wavelength $\lambda$, so that $w(z)\approx w_0$ and remains approximately constant over a propagation distance of many beam diameters. 

The electrostrictive force due to the presence of the beam is $\nabla \chi$ where
\be
  \chi(\rho,t) = \half \varepsilon_0 E^2(\rho,t) \varrho_m \frac{\rmd n^2}{\rmd \varrho_m} = \chi(0)T(t)\exp\left[-\frac{2\rho^2}{w^2(z)}\right]\label{4}
\ee
where $\varrho_m$ is the fluid's density and $\chi(0)$ is the initial and original value of $\chi$. The speed of light is much greater than that of sound, hence we can assume all changes in the electromagnetic laser field to be instantaneous and simultaneous throughout the liquid. We then clearly have $\chi(\rho,t)=\chi(\rho)T(t)$. Assuming the liquid to obey the one-component Lorentz-Lorenz (Clausius-Mossotti) relation
\be\label{clausius-mossotti}
  \frac{\varrho_m \alpha}{\varepsilon_0} = \frac{n^2-1}{n^2+2}
\ee
where $\alpha$ is molecular polarizability gives
\be
  \chi(\rho) = \sixth \varepsilon_0 E^2(\rho) (n^2-1)(n^2+2).
\ee
Explicitly,
\be
  \chi(0) = \frac{(n^2-1)(n^2+2)P}{3\pi n c w_0^2}.
\ee
It is to be noted that the Clausius-Mossotti relation, according
to which the permittivity depends on the fluid density only,
refers to a non-polar fluid. The relation is convenient to use in
practice, as one does not have to distinguish between the
isothermal derivative $(\partial n^2/\partial \varrho_m)_T$ and
the adiabatic derivative $(\partial n^2 /\partial
\varrho_m)_S$. It is found to apply with good accuracy for
gases, for many liquids, and even for some solids. For polar
liquids, in order to get better accuracy, the
Onsager relation is available \cite{onsager36} with various modifications
\cite{bottcher73}. In our theory the use of another relation for $n(\varrho_m)$ amounts only to a different form of $\chi(0)$, and since we will work to linear order in the density variations in the liquid, $\chi(0)$ will enter only as a prefactor throughout the below theory. Hence the choice of relation $n(\varrho_m)$ does not affect the following calculations and results in any essential way.

We ought to point out at this point that we ignore other effects besides electrostriction, in particular, the quadratic electro-optic effect. It is known that in strong electric fields a liquid can become optically anisotropic, the polarizability $\alpha_\parallel$ parallel to the field being different from the polarizability $\alpha_\perp$ in the transverse direction. The field dependence of the polarizability in turn influences the refractive index through the Lorentz-Lorenz relation (the optical analogue of the Clausius-Mossotti relation).  On may estimate the magnitude of the quadratic electro-optic effect by considering the nonlinear quantum theory of the refractive index in an isotropic medium. 
For instance, regarding an artificial medium consisting of two-level atoms (see p.\ 147 of Ref.~\onlinecite{brevik79} and further references therein), it follows that $\alpha_\parallel =\alpha_0/(1+E^2/E_0^2)$. Here $\alpha_0$ is the polarizability in the absence of the field and $E_0$ is the constant $E_0=\hbar \omega_s/(\sqrt{2}e x_{0s})$, where $\hbar \omega_s$ is the energy difference between the two energy levels of the atom and $x_{0s}$ the matrix element of the position operator between the same levels. For simple non-polar liquids, typically $\omega_s \sim 10^{16}\,\rm{s}^{-1}$, so that if the matrix element is in the order of the Bohr radius, we obtain $E_0 \sim 10^{10}$ to $10^{11}$ V/m.  This is a very high value, implying that the quadratic electro-optic effect can be safely ignored under usual physical conditions.

To study the pressure that builds due to the electrostrictive force, consider the linearized Euler equation
\be\label{euler}
  \varrho_m \partial_t \bd{v} = -\nabla p + \nabla \chi
\ee
[$\partial_t \equiv \partial/\partial t$]. It follows from \eqref{euler} and the fact that the same equation must hold before the arrival of the beam, that $\nabla\times \bd{v}=0$, and hence we may write the velocity field in the form of a velocity potential,
\be
  \bd{v}=\nabla \Phi.
\ee
Thus, because of gauge invariance of potentials we get the simple equation
\be
  p(\rho,t) = -\varrho_m \partial_t \Phi(\rho,t) + \chi(\rho,t).
\ee
Writing $p(\rho,t)=p_0 + p'(\rho,t)$ and $\varrho_m(\rho,t)=\varrho_{m0} + \varrho_m'(r,t)$ where $p_0, \varrho_{m0}$ are the constant, initial pressure and density, and $p'\ll p_0$, $\varrho_m'\ll \varrho_{m0}$, we linearize w.r.t.\ $p', \varrho_m'$ and go to the gauge frame,
\be\label{pphi}
  p'(\rho,t) = -\varrho_{m0} \partial_t \Phi(\rho,t) + \chi(\rho,t).
\ee
The second term on the right hand side is due to the direct electrostrictive force from the local electric field density whereas the first term comes from propagation of pressure waves from the surrounding area. Spherical pressure waves are emitted from all points within the beam when the electric field varies with time. [Field variations over an optical period is of no consequence; the material response time is much longer than this, and these periodic field variations simply average to zero.] With the continuity equation
\be
  \partial_t \varrho_m + \nabla\cdot(\varrho_m \bd{v}) = 0
\ee
and noting that
\be
  p' = u^2 \varrho_m'
\ee
where $u = [(\partial p'/\partial \varrho_m)_S]^{1/2}$ is the speed of sound, the governing equation for $\Phi(\rho,t)$ is obtained
\be\label{chieq}
  (\nabla^2 - u^{-2}\partial_t^2)\Phi(\rho,t) = -\frac1{\varrho_{m0} u^2} \partial_t \chi(\rho,t).
\ee

\section{Step-function switch-on of beam}\label{sec_step}

Assume first that the beam is switched on suddenly at $t=0$ and off again at $t_0$.
\be\label{Tstep}
  T(t) = \Theta(t) - \Theta(t-t_0)
\ee
where $\Theta$ is the unit step function. Because the velocity of light is immensely greater than the velocity of sound, we may safely assume the onset of electrostrictive forces to be simultaneous throughout the liquid. We have
\be
  \partial_t \chi(\rho,t) = \chi(\rho)\partial_t T(t)=\chi(\rho)[\delta(t)-\delta(t-t_0)].
\ee
We define the Green's function of Eq.~\eqref{chieq}, which by definition satisfies
\be
  (\nabla^2 - u^{-2}\partial_t^2)G(\bd{r},\bd{r}',t,t') = \delta^3(\bd{r}'-\bd{r}) \delta(t'-t)
\ee
which has solution (e.g.\ Ref.~\onlinecite{jackson99} p.~244)
\be\label{green}
  G(\bd{r}',\bd{r},t,t') = -\frac1{4\pi|\bd{r}'-\bd{r}|}\delta\left(t'-t+\frac{|\bd{r}'-\bd{r}|}{u}\right).
\ee
This solution requires that scattering of sound waves off fluid surfaces may be neglected. In microscopic systems where such effects are of importance, a Green's function taking these boundaries into account is required -- Green's functions for the scalar wave equation have been found for a wealth of geometries.

From \eqref{chieq} and \eqref{green} it follows that (noting $\delta(ax)=\delta(x)/|a|$)
\begin{align}
  \Phi&(\rho,t) \notag \\
  =& -\frac1{\varrho_{m0} u^2}\int_{-\infty}^t \rmd t' \int \rmd^3 r' G(\bd{r}',\bd{r},t,t') \partial_t \chi(r',t') \label{greensfunction} \\
  =& \frac{\chi(0)}{\pi\varrho_{m0} u}\int_0^{\pi} \rmd \theta' \int_0^\infty \rmd z' \int_0^\infty \rho' \rmd \rho' \frac{\exp(-2\rho^{\prime 2}/w^2_0)}{R}\notag \\
  &\times [\delta(tu-R)-\delta(ut-ut_0 -R)]\label{phidelta}
\end{align}
where we define
\be
  \bd{R} = |\bd{r}'-\bd{r}|; ~~ R = |\bd{R}|.
\ee
Since the geometry is obviously symmetric under $z\to -z$ and $\theta\to -\theta$ we have halved the integration ranges of $z'$ and $\theta'$ above. For the same reason we are free to choose $\bd{r}$ to lie on the $x$ axis without loss of generality: $z=0$, $\theta=0$. To integrate out the remaining delta function we integrate with respect to $R$ using
\be
  R^2 = \rho^2 + \rho'^2 - 2\rho\rho'\cos\theta' + z^{\prime 2}
\ee
to substitute 
\be
  \frac{\rmd z'}{R} = \frac{\rmd R}{z'} = \frac{\rmd R }{\sqrt{R^2 - \rho^2 - \rho'^2 + 2\rho\rho'\cos\theta'}}.
\ee
The delta functions in Eq.~\eqref{phidelta} pick out two spherical shells of values of $(r',\theta')$ centered at $\bd{r}$ and having radii $ut$ and, for $t>t_0$, $u(t-t_0)$, respectively. 
Thus, considering the term of radius $ut$ [the second term is essentially identical, as is clear from Eq.~(\ref{phidelta})]
\begin{align}
  &\int_0^{\pi} \rmd \theta' \int_0^\infty \rho' \rmd \rho'f(\rho')\int_0^\infty \frac{\rmd z'}{R}\delta(tu-R)\notag \\
  &= \int_0^{\pi} \rmd \theta' \int_0^\infty \rho' \rmd \rho'f(\rho') \int_{\mathcal R}^\infty \frac{\rmd R~ \delta(R-tu)}{\sqrt{R^2 - \mathcal R^2}}\notag \\
    &= \int_0^\pi\rmd \theta' \int_0^\infty\frac{\rho' \rmd \rho'f(\rho')}{\sqrt{u^2 t^2 - \mathcal R^2}}\Theta(ut-\mathcal R)
\end{align}
where we define
\be
   \mathcal R^2(t) = \rho^2 + \rho'^2 - 2\rho\rho'\cos\theta'.
\ee
In the above, the exponential in (\ref{phidelta}) depending only on $\rho'$ were gathered in a shorthand function $f(\rho')$.

\begin{figure}[htb]
  \begin{center}
    \includegraphics[width=2.6in]{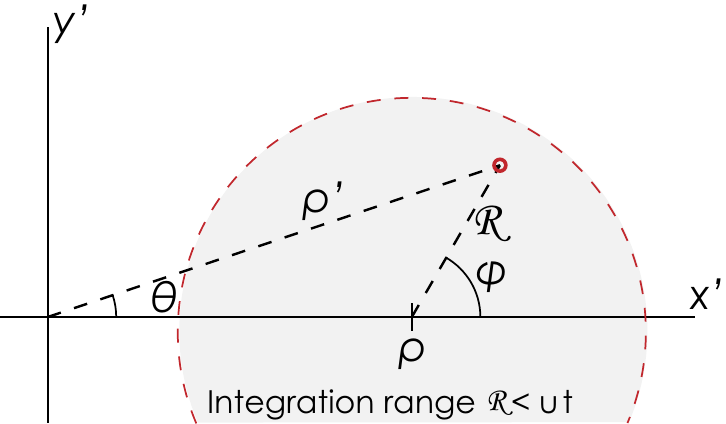}
    \caption{Coordinate shift: translating the $z$ axis to the point $(\rho,0)$.}
    \label{fig:coordtrans}
  \end{center}
\end{figure}

The $\Theta$ function restricts the integration area to a circular disc of radius $ut$ centered at $(\rho,0,0)$. It is natural to integrate instead by parametrising according to a cylidrical coordinate system where the center of this disc lies at the origin as shown in figure \ref{fig:coordtrans}. We note that in a new coordinate system $(\mathcal R, \varphi)$,
\be
  \rho^{\prime 2} = \rho^2 + \mathcal R^2 + 2 \rho \mathcal R \cos\varphi.
\ee
Now the $\Theta$ function is equivalent with the upper integration limits, and can be removed. An equivalent expression is obtained for the second term, but with $t\to t-t_0$, and only allowing contributions for $t>t_0$ for reasons of causality.

Using the relation
\be
  \int_0^\pi \frac{\rmd \varphi}{\pi} \rme^{-a \cos\varphi} = I_0(a)
\ee
where $I_0(x)$ is the modified Bessel function of the first kind of order zero, we thus obtain a general expression for $\Phi(\rho,t)$:
\begin{align}
  &\Phi(\rho,t) = \frac{\chi(0)}{\varrho_{m0} u}\left\{ \Theta(t)\int_{0}^{ut}\mathcal R\rmd \mathcal R \frac{I_0\left(\frac{4\rho\mathcal R}{w_0^2}\right)\exp\left(-\frac{2(\rho^2 + \mathcal R^2)}{w_0^2}\right)}{\sqrt{u^2t^2-\mathcal R^2}}  \right.\notag \\
  &-\left.\Theta(t-t_0)\int_0^{u(t-t_0)}\!\!\!\!\!\!\mathcal R\rmd \mathcal R
  \frac{I_0\left(\frac{4\rho\mathcal R}{w_0^2}\right) 
  \exp\left(-\frac{2(\rho^2 + \mathcal R^2)}{w_0^2}\right)}{\sqrt{u^2(t-t_0)^2-\mathcal R^2}}
  \right\} \label{bigequation}
\end{align}
The gauge pressure can now be calculated Eq.~\eqref{pphi}, and we shall do so in a moment. First, we will simplify our notation considerably by introducing rescaled variables for time and length.

\subsection{Rescaled time and radius}

For further analysis we introduce the typical time and length scales
\be
  \tilde \rho = \frac{w_0}{\sqrt{2}};~~   \tilde t = \frac{w_0}{\sqrt{2}u}
\ee
satisfying $u\tilde t = \tilde \rho$. The corresponding dimensionless time and space variables we define as
\begin{subequations}\label{dimless}
\begin{align}
  \xi^{\prime} = \rho^{\prime}/\tilde \rho; ~~~ \xi = \rho/\tilde \rho;\\
   \tau =t/\tilde t; ~~~ \tau_{0} =t_{0}/\tilde t; 
\end{align}
\end{subequations}
and substituting for $\mathcal R\to x=
   x = \sqrt{u^2t^2-\mathcal R^2}/\tilde \rho$ as well as rescaling we get the much simpler looking general expressions
\begin{subequations}
  \begin{align}
    \Phi(\rho,t) =& \frac{\chi(0) \tilde t}{\varrho_{m0} }\left[ \tilde \Phi(\xi, \tau)-\tilde \Phi(\xi,\tau-\tau_0)\right];\\
    \tilde \Phi(\xi, \tau) =&  \Theta(\tau)\rme^{-\xi^2} \int_0^{\tau} \rmd x I_0(2\xi\sqrt{\tau^2-x^2})\rme^{x^2-\tau^2}.\label{tildephistep}
  \end{align}
\end{subequations}
The integral in Eq~(\ref{tildephistep}) is simple to evaluate numerically. With the re-scaled variables and the dimensionless $\tilde \Phi$ function defined above, Eq.~\eqref{pphi} becomes [the re-scaled time envelope function is obviously $T(\tau)=\Theta(\tau)-\Theta(\tau-\tau_0)$]
\begin{align}
  p(\xi,\tau) =& -\chi(0)[\partial_\tau \tilde \Phi(\xi, \tau)-\partial_\tau\tilde \Phi(\xi,\tau-\tau_0)\notag \\
  &- T(\tau)\rme^{-\xi^2}].\label{pphirescaled}
\end{align}
Evaluating the derivatives we finally find the pressure distribution as function of (rescaled) radius and time
\begin{subequations}\label{pstep}
  \begin{align}
    p(\xi,\tau) =& \chi(0)[P(\xi,\tau)-P(\xi,\tau-\tau_0)]\label{Pdef} \\
    P(\xi,\tau) =& 2\tau\rme^{-\xi^2}\Theta(\tau)\int_0^\tau \frac{x\rmd x \rme^{-x^2}}{\sqrt{\tau^2-x^2}}\Bigl[ I_0(2\xi x)\notag \\
    &-\frac{\xi}{x}I_1(2\xi x)\Bigr]\label{pressureStep}
  \end{align}
\end{subequations}
where in the last line we have substituted $x\to \sqrt{\tau^2-x^2}$. Equations (\ref{pstep}) are the final result of the pressure calculation in the case of step-function switch-on and -off of the laser.

\begin{figure*}[htb]
  \begin{center}
    \includegraphics[width=\textwidth]{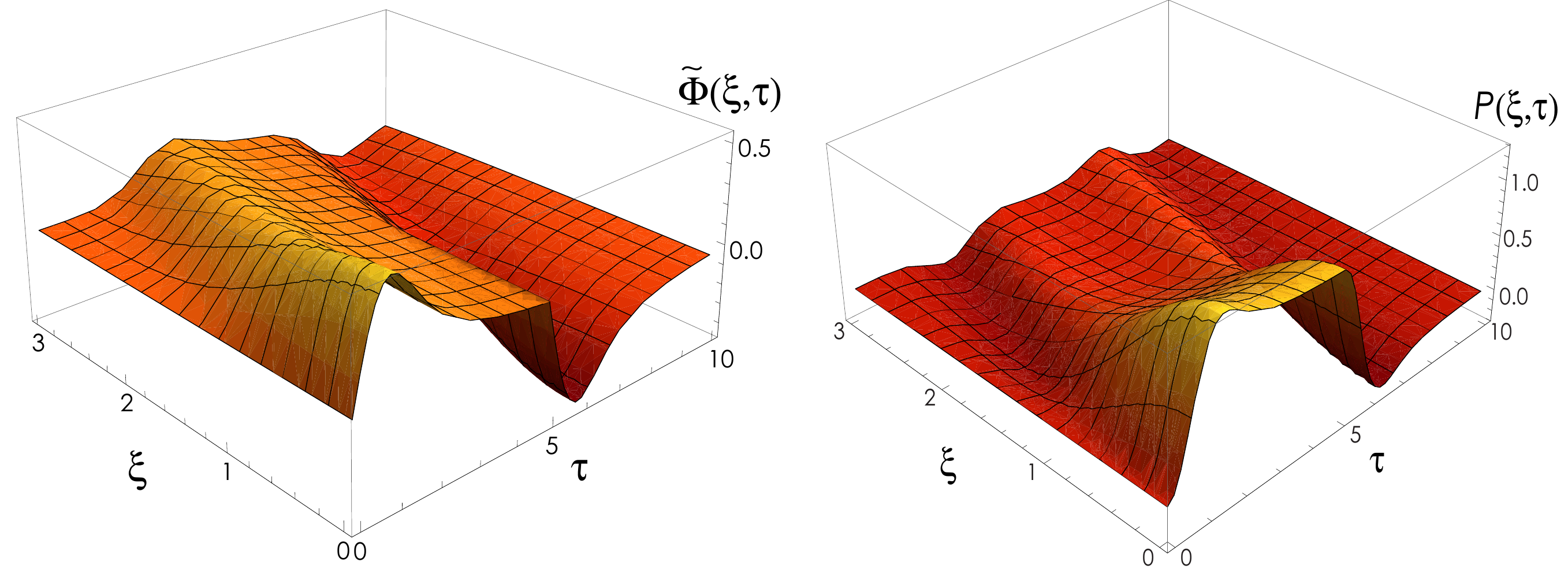}
    \caption{Left: the function $\tilde \Phi(\xi,\tau)$ for $\tau_0=5$. Right: The reduced pressure $p(\xi,\tau)/\chi(0)$ for $\tau_0=5$ as given in Eq.~\eqref{pressureStep}.}
    \label{fig:phipstep}
  \end{center}
\end{figure*}

The generalisation to a series of $N$ pulses turned on at times $t_i$ and off at times $t_{0i}$ ($t_i<t_{0i}<t_{i+1}$) is obvious:
\begin{align}
  p(\xi,\tau) =& \chi(0)\sum_{i=1}^N [P(\xi,\tau_i) - P(\xi,\tau-\tau_{0i})].
\end{align}

A plot of the functions $\tilde\Phi(\xi,\tau)$ and $P(\xi,\tau)$ are shown in figure \ref{fig:phipstep}. One notes how the pressure on the axis rises steeply when the laser is switched on and then relaxes monotonously towards a new equilibrium value which has the very simple form
\be\label{asymptotic}
  P(\xi,\tau) \buildrel{\tau\gg 1}\over{\longrightarrow} \rme^{-\xi^2}.
\ee
We show this in the appendix. While it is not entirely trivial to obtain the limit from the more general expression, this result could have been obtained in a very straightforward way by assuming a static laser beam from the beginning. After reaching the peak value, the pressure relaxes quickly and the stationary situation is reached already at around $3\tilde t$. The left panel of figure \ref{fig:phipstep} also shows how a pressure wave travels outwards from the cylinder axis after switch-on at velocity $u$, and similarly, a wave of negative gauge pressure at switch-off. This fact allows closely controlled acoustic waves to be created by a laser beam, the principle behind laser-induced thermal acoustics, already used for measurement purposes\cite{cummings94,cummings95}.

\subsection{Pressure time variations on the axis}

\begin{figure}[htb]
  \begin{center}
    \includegraphics[width=3.2in]{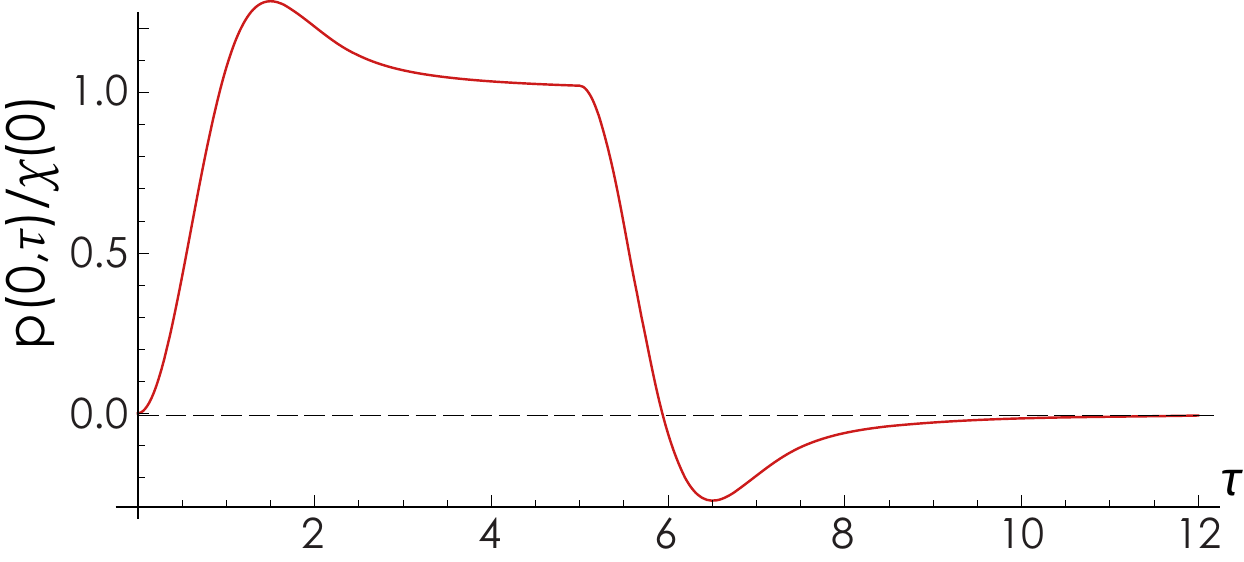}
    \caption{The pressure $p(0,\tau)$ on the axis, Eq.~\eqref{pressureAxis}, as function of dimensionless time $\tau$, divided by the prefactor $\chi(0)$. The pulse is turned on at $\tau=0$ and off at $\tau=\tau_0=5$.}
    \label{fig:onaxisStep}
  \end{center}
\end{figure}

A simple analytical expression can be found for points on the $z$ axis, $\xi=0$. Then we are left with
\be
  \tilde\Phi(0,\tau)=\Theta(\tau)\int_0^\tau\rmd x \exp(x^2-\tau^2) = \Theta(\tau)F(\tau)
\ee
where $F(x)=\rme^{-x^2}\int_0^x\rmd s \rme^{s^2}$ is Dawson's integral.
Inserting this into Eq.~\eqref{pphirescaled} with $F'(x)=1-2xF(x)$ we get simply, in terms of the dimensionless units Eq.~\eqref{dimless},
\begin{align}
  \tilde\Phi(0,\tau) =& \Theta(\tau)F(\tau)-\Theta(\tau-\tau_0)F(\tau-\tau_0);\\
  p(0,\tau) =& 2\chi(0)[\Theta(\tau)\tau F(\tau)\notag \\
  &-\Theta(\tau-\tau_0)(\tau-\tau_0)F(\tau-\tau_0)].\label{pressureAxis}
\end{align}
The dimensionless pressure $p(0,\tau)/\chi(0)$ is shown in figure \ref{fig:onaxisStep}. The special case of the cylinder axis was considered also previously\cite{brevik79}, in agreement with Eq.~(pressureAxis).

\section{Laser pulse with general time evolution}

Whereas we considered a sharp turn-on and turn-off of the beam in the previous section, here we will be concerned with a more typical pulse transient. Within the same formalism as employed above, we are able to keep the function $T(t)$ general throughout, yielding results of more general validity. For numerical examples in the following, we choose the following typical form for a soft laser pulse,
\be\label{softtransient}
  T(t) = \frac{\rme^2}{4}\frac{\tau^2}{\tau_0^2}\rme^{-\tau/\tau_0}
\ee
so that $t_0=\tilde t \tau_0$ is again the duration of the pulse. Here and forthwith we make use of the rescaled variables defined in Eq.~\ref{dimless}. We have
\begin{align}
  T'(\tau) =&
        \frac{\rme^2}{4\tau_0}\left(2\frac{\tau}{\tau_0}-\frac{\tau^2}{\tau_0^2}\right)\rme^{-\tau/\tau_0};\\
  T''(\tau) =&
        \frac{\rme^2}{4\tau_0^2}\left(2-4\frac{\tau}{\tau_0}+\frac{\tau^2}{\tau_0^2}\right)\rme^{-\tau/\tau_0}.\label{Tpp}
\end{align}
The velocity potential is again given by Eq.~\eqref{greensfunction}, wherewith we get after rescaling
\begin{align}
  \tilde \Phi(\xi,\tau)=&\frac1{4\pi}\int_0^\tau \rmd \tau' \int \rmd {\boldsymbol \xi}' \frac{\rme^{-\xi^{\prime 2}}}{s}\delta(\tau'-\tau+s) T'(\tau')\notag \\
  =&\frac1{4\pi}\int_{s\leq \tau} \rmd {\boldsymbol \xi}' \frac{\rme^{-\xi^{\prime 2}}}{s} T'(\tau-s)
\end{align}
where we define
\be
  \mathbf{s} = {\boldsymbol \xi}'-{\boldsymbol \xi};~~ s = |\mathbf{s}|.
\ee

\begin{figure}[htb]
  \begin{center}
    \includegraphics[width=2.7in]{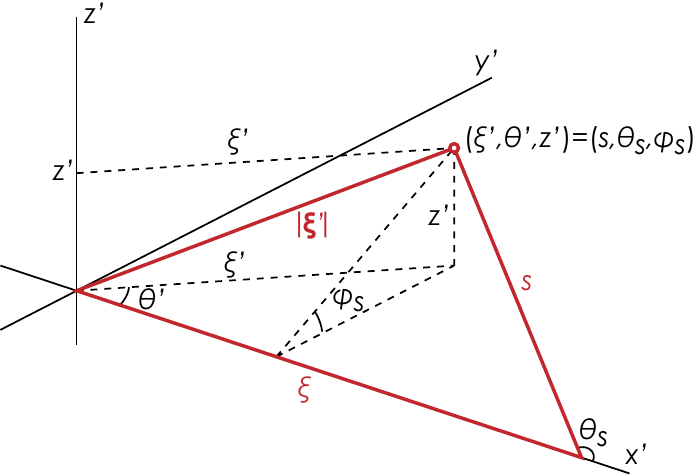}
    \caption{Coordinate transformation to spherical coordinates with origin at $(\xi,0,0)$ in the cylindrical $(\xi',\theta',z')$ system, and polar axis $x'$.}
    \label{fig:spherical}
  \end{center}
\end{figure}

As before, we are free to choose ${\boldsymbol \xi} = (\xi,0,0),$ and note that the integrand is invariant under rotation about the $\xi'_x$ axis. The integral over $\xi'$ runs over all of space, so we choose integration coordinates instead to be spherical coordinates with the origin at ${\boldsymbol \xi}$ in the ${\boldsymbol \xi}'$ system and with the $\xi'_x$ axis as polar axis with polar angle $\theta_s$. In the new coordinate system $(s,\theta_s,\varphi_s)$ the integral reads
\begin{align}
  \tilde \Phi&(\xi,\tau)=\frac1{4\pi}\int_0^\tau s \rmd sT'(\tau-s)\int_0^{2\pi}\rmd \varphi_s  \int_0^\pi \sin\theta_s \rmd \theta_s  \notag \\
  &\times\exp[-\xi^2-s^2(1- \sin^2\theta_s \sin^2\varphi_s)-2\xi s \cos\theta_s]
\end{align}
where the exponent is $-\xi^{\prime 2}(\xi,s,\theta_s,\phi_s)$, as can be seen geometrically from figure \ref{fig:spherical} when noting that the length $z'$ can be expressed as $z'=s\sin\theta_s\sin\phi_s$ and that $|{\boldsymbol\xi}'|^2=\xi^2+s^2+2\xi s\cos\theta_s=\xi^{\prime 2}+z^{\prime 2}$. 

We now make use of the relation
\[
  \int_0^{2\pi} \rmd \phi_s \rme^{s^2 \sin^2\theta_s \sin^2\phi_s} = 2\pi\rme^{\frac1{2}s^2 \sin^2\theta_s}I_0(\half s^2 \sin^2\theta_s)
\]
giving
\begin{align}
  \tilde \Phi(\xi,\tau)
    =&\rme^{-\xi^2}\int_0^\tau  \rmd s T'(\tau-s)\rme^{-\frac{s^2}{2}}\notag\\
    &\times\int_0^s \rmd \beta \rme^{-\frac1{2} \beta^2}I_0[\half(s^2- \beta^2)]  \cosh 2\xi \beta \label{PhiPulse}
\end{align}
where we substituted $\beta=s\cos\theta_s$.

As a consistency check we should be able to insert the step-function time evolution derivative of Eq.~\eqref{Tstep}, $T'(\tau)=\delta(\tau)-\delta(\tau-\tau_0)$ and get back the expression Eq.~\eqref{tildephistep}. One quickly sees that this is so, provided
\begin{align}
  \int_0^\tau &\rmd x \rme^{x^2} I_0(2\xi \sqrt{\tau^2-x^2}) \notag \\
  &= \int_0^\tau \rmd \beta \rme^{\half(\tau^2-\beta^2)}I_0[\half(\tau^2-\beta^2)] \cosh 2\xi \beta\label{ansatz}
\end{align}
for arbitrary $\tau$ and $\xi$. We have verified the identity numerically in the positive-positive $(\tau,\xi)$ plane. The two different derivations of the re-scaled velocity potential for the step-function pulse thus serve as derivation of the potentially useful relation Eq.~(\ref{ansatz}).

\begin{figure}[htb]
  \begin{center}
    \includegraphics[width=2.5in]{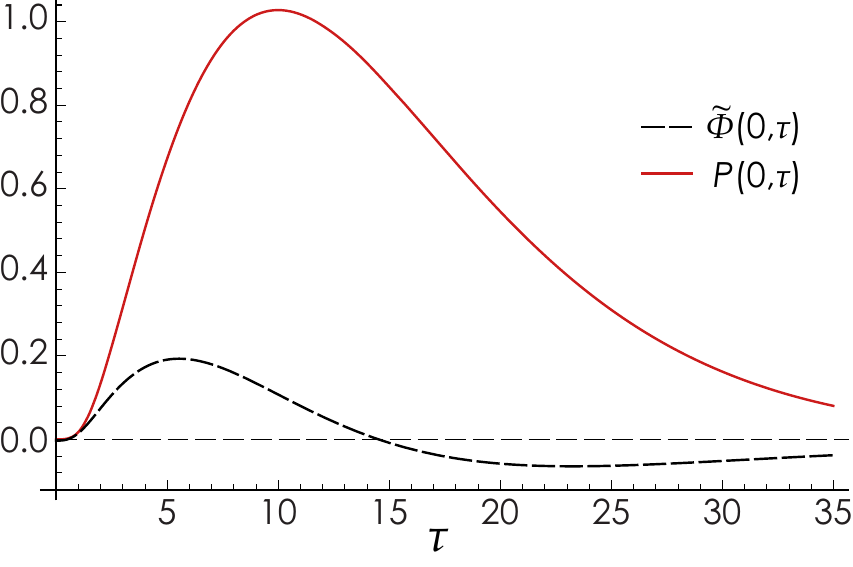}
    \caption{The functions $\tilde \Phi$ and $P$ on the symmetry axis for $\tau_0=5$.}
    \label{fig:phipaxis}
  \end{center}
\end{figure}

As in the previous case, the expression can be simplified on the axis. Let us first use Eq.~\eqref{ansatz} to write Eq.~\eqref{PhiPulse} as
\begin{align}
  \tilde \Phi(\xi,\tau)=&\rme^{-\xi^2}\int_0^\tau  \rmd s T'(\tau-s)\rme^{-s^2}\notag \\
  &\times\int_0^s \rmd x \rme^{x^2} I_0(2\xi \sqrt{s^2-x^2}),\label{PhiPulse2}
\end{align}
an expression which is still valid for an arbitrary $T(\tau)$. On the $z$ axis ($\xi=0$) $\tilde\Phi$ becomes simply
\be
  \tilde\Phi(0,\tau) =\int_0^\tau \rmd sT'(\tau-s)F(s)
\ee
where $F(x)$ is again Dawson's integral. The functions $\tilde \Phi(0,\tau)$ and $P(0,\tau)$ are plotted in figure \ref{fig:phipaxis}.

The expression for the pressure is simple to write down from the generalization of Eq.~\eqref{pphirescaled},
\be
  p'(\xi,\tau)=\chi(0)[T(\tau)\rme^{-\xi^2}-\partial_\tau \tilde\Phi(\xi,\tau)].
\ee
Analogous to and for comparison with Eq.~\eqref{Pdef}, we define
\be
  p(\xi,\tau) = \chi(0)P(\xi,\tau)
\ee
now with
\begin{align}
  P&(\xi,\tau) = \rme^{-\xi^2}\Biggl[T(\tau)\notag \\
  &-\int_0^\tau  \rmd s T''(\tau-s)\rme^{-s^2}\int_0^s \rmd x \rme^{x^2} I_0(2\xi \sqrt{s^2-x^2}) \notag \\
  &-T'(0)\rme^{-\tau^2}\int_0^\tau \rmd x \rme^{x^2} I_0(2\xi \sqrt{\tau^2-x^2}) \Biggr].\label{Ppulse}
\end{align}
This the expression for the pressure given a general $T(\tau)$, although this form clearly requires the existence of $T'(0)$ which is not satisfied for the step function considered in section \ref{sec_step}. For the soft pulse transient of Eq.~(\ref{softtransient}), $T'(0)=0$ and the last term vanishes.

\begin{figure*}[htb]
  \begin{center}
    \includegraphics[width=\textwidth]{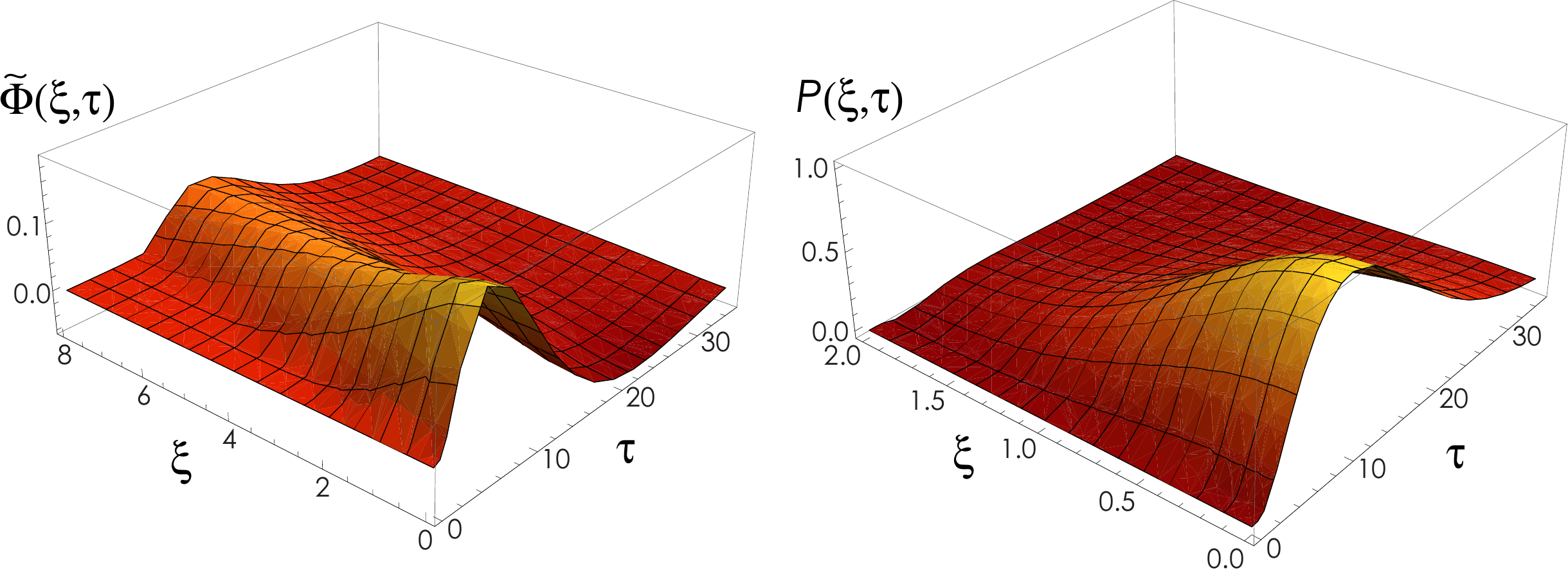} 
    \caption{Left: the function $\tilde \Phi(\xi,\tau)$ from Eq.~\eqref{PhiPulse} or \eqref{PhiPulse2} for $\tau_0=5$. Right: The reduced pressure $p(\xi,\tau)/\chi(0)$ for $\tau_0=5$ as given in Eq.~\eqref{Ppulse}.}
    \label{fig:phipPulse}
  \end{center}
\end{figure*}

We plot $\tilde\Phi(\xi,\tau)$ and $P(\xi,\tau)$ using the pulse transient (\ref{softtransient}) in figure \ref{fig:phipPulse}. Notice how much longer it takes for the gauge pressure to relax to zero again compared to the step function turn-off shown in figure \ref{fig:phipstep}. The behaviour is otherwise comparable to that of figure \ref{fig:phipstep}. The generality of Eq.~(\ref{Ppulse}) implies for example that one can derive the exact acoustic pressure wave field throughout the medium which is set up by a train of laser pulses with known time dependence [reflection/absorption at material boundaries not included, of course -- this would require a different, geometry specific Green's function in Eq~(\ref{greensfunction})]. 

Also in this case the expression takes a simpler form on the symmetry axis, where we are left with
\be
  P(0,\tau)=T(\tau)-T'(0)F(\tau) - \int_0^\tau \rmd s T''(\tau-s)F(s)
\ee
where $F(s)$ is again the Dawson integral.

\section{Conclusions}

We have calculated the local gauge pressure distribution due to electrostrictive forces as a function of time and position, in the presence of a Gaussian laser beam propagating through a bulk fluid. The resulting expressions are simple and may form a natural part of an analytical toolkit for experimental interpretation. A typical scale for length ($\tilde\rho$) and time ($\tilde t$) emerge, given by the beam width and the velocity of sound. Our final results are valid for a general time evolusion of the propagating laser beam, and for demonstratino purposes we consider two typical cases: a sudden switch-on/switch-off of the beam power, as well as a soft laser pulse. Reasonably simple expressions are obtained for both cases, as given, respectively, by our Eqs.~\eqref{pstep} and \eqref{Ppulse} in terms of the re-scaled distance from the beam axis ($\xi=\rho/\tilde\rho$), and time $(\tau=t/\tilde t)$. On the beam axis, these expressions may be simplified further. The pressure relaxation time, the duration before the gauge pressure has returned to zero as the laser pulse is turned off, is surprisingly different in the two cases for the same pulse duration. 

Provided a stationary laser is switched on from zero to full intensity in a time shorter than the typical time scale $\tilde t = w_0/\sqrt{2}u$ ($w_0$: beam waist, $u$: speed of sound), our result for a step function switch on are valid. After a transient period in which the pressure rises sharply and then relaxes to a stationary value, equilibrium is reached after a time of $3-4\tilde t$, roughly the time for a sound wave to propagate the beam diameter. Under most circumstances fluid motion is much slower than this, and electrostrictive pressure can be considered to be set up instantaneously. A cylindrical acoustic pressure wave travels outwards from the beam on switch-on, and similarly a wave of lower pressure is emitted at switch-off, allowing for highly controlled laser-induced acoustics in the medium, a principle already used in measurement set-ups\cite{cummings94,cummings95}. Our result \eqref{Ppulse} provides in a simple way the acoustic pressure field which results from an arbitrary train of laser pulses anywhere in the medium at any time.

Optofluidics and electrohydrodynamics are fields of research seeing impressive advances and promises many applications in optics, sensing and measurement, microfluidics and microchemistry, see e.g.\ Refs.~\onlinecite{monat07,garnier03} and references therein. So far this field of research has been driven by impressive experimental advances with theorists working to explain the observed phenomena. In this paper we provide another theoretical building block towards this end. Electrostriction, while not typically required to describe electromagnetically induced fluid deformations and flows themselves, is often of importance to their stability. This is examplified by perhaps the simplest example of electrohydrodynamics, that of a liquid rising between condensator plates due to the electric field. While the elevation can be calculated without reference to electrostriction, the latter plays the tacit but crucial role of enabling the liquid column to rise as a coherent whole\cite{brevik79,brevik81}. The methods and results presented herein could have many applications in analyses of laser driven flow and fluid manipulation, and for the stability of these.

\appendix
\section{Equilibrium value after step-function switch-on}

We show here Eq.~(asymptotic), the asymptotic equilibrium value at times $\tau\gg 1$ long after switch-on of a step-function laser, assuming no switch-off has occured. For $\tau\gg 1$, Eq.~(\ref{pressureStep}) becomes
\begin{align}\label{Plimit}
  P(\xi,\tau) \to 2\rme^{-\xi^2}\int_0^\tau \rmd x x\rme^{-x^2}\Bigl[ I_0(2\xi x)-\frac{\xi}{x}I_1(2\xi x)\Bigr]
\end{align}
Using the following formulae from Ref.~\onlinecite{Gradshteyn80}:
\begin{align*}
  \int_0^\infty \rmd xx^{\nu+1}\rme^{-\alpha x^2}J_\nu(\beta x) =& \frac{\beta^2}{(2\alpha)^{\nu+1}}\rme^{-\beta^2/4\alpha}\\
  \int_0^\infty \rmd xx^{\nu-1}\rme^{-\alpha x^2}J_\nu(\beta x) =& 2^{\nu-1}\beta^{-\nu}\gamma(\nu,\beta^2/4\alpha)
\end{align*}
where $\gamma(a,x)=\int_0^x \rmd t \rme^{-t}t^{a-1}$ is the incomplete Gamma function, 
using $J_n(\rmi z)=i^nI_n(z)$ and letting $\beta=2\rmi \xi$ we find
\begin{align}
  \int_0^\infty \rmd x x\rme^{-x^2} I_0(2\xi x) =& \frac1{2}\rme^{\xi^2},\\
  \int_0^\infty \rmd  x\rme^{-x^2} I_1(2\xi x) =& -\frac1{2\xi}(1-\rme^{\xi^2}).
\end{align}
Inserted into (\ref{Plimit}) this immediately gives Eq.~(\ref{asymptotic}).

\end{document}